\documentclass[useAMS,usenatbib]{mn2e}
\usepackage{graphicx}
\usepackage{subfigure}

\newcommand{\Msun}{\mbox{$M_{\odot}$}}
\newcommand{\degree}{^{\scriptscriptstyle{\circ}}}

\title[X-ray Polarization in Jets]{X-ray Polarization in Relativistic Jets}
\author[McNamara et al.]
{Aimee L. McNamara$^{1}$\thanks{E-mail: aimee@physics.usyd.edu.au},
Zdenka Kuncic$^{1}$ 
and Kinwah Wu$^{2,3}$  \\
$^{1}$School of Physics, University of Sydney, NSW 2006, Australia\\
$^{2}$Mullard Space Science Laboratory, University College London, Holmbury St Mary, Surrey, RH5 6NT \\ 
$^{3}$Department of Physics, University of Hong Kong, Pokfulam Road, Hong Kong SAR}  
\begin{document}

\date{Accepted . Received ; in original form }

\pagerange{\pageref{firstpage}--\pageref{lastpage}} \pubyear{2008}

\maketitle

\label{firstpage}

\begin{abstract}
We investigate the polarization properties of Comptonized X-rays from relativistic jets in Active Galactic Nuclei (AGN) using Monte Carlo simulations. We consider three scenarios commonly proposed for the observed X-ray emission in AGN: Compton scattering of blackbody photons emitted from an accretion disk; scattering of cosmic microwave background (CMB) photons; and self-Comptonization of intrinsically polarized synchrotron photons emitted by jet electrons. Our simulations show that for Comptonization of disk and CMB photons, the degree of polarization of the scattered photons increases with the viewing inclination angle with respect to the jet axis. In both cases the maximum linear polarization is $\approx 20\%$. In the case of synchrotron self-Comptonization (SSC), we find that the resulting X-ray polarization depends strongly on the seed synchrotron photon injection site, with typical fractional polarizations $P \approx 10 - 20\%$ when synchrotron emission is localized near the jet base, while $P\approx 20 - 70\%$ for the case of uniform emission throughout the jet. These results indicate that X-ray polarimetry may be capable of providing unique clues to identify the location of particle acceleration sites in relativistic jets. In particular, if synchrotron photons are emitted quasi-uniformly throughout a jet, then  the observed degree of X-ray polarization may be sufficiently different for each of the competing X-ray emission mechanisms (synchrotron, SSC or external Comptonization) to determine which is the dominant process. 
However, X-ray polarimetry alone is unlikely to be able to distinguish between disk and CMB Comptonization.

\end{abstract}


\section{Introduction}
Current X-ray observations are limited to spectral characteristics and timing variability over a relatively narrow electromagnetic window ($\sim 0.1 - 50$ keV). This often leads to ambiguities in the interpretation of the data, where two or more different models can explain the same observations with equal success. This degeneracy may be resolved if polarization measurements are available. X-ray Polarization observations may provide more precise diagnostic information about the emission processes and the geometries of the emission regions \citep[see e.g.][]{ST85,Lapidus85}. Considerable effort is currently underway to develop the X-ray instruments capable of measuring polarization of at least a few percent \citep{Costa01,Costa08}.  
  
The last dedicated experiment to measure X-ray polarization of astronomical sources beyond the solar system was conducted over thirty years ago \citep[see][for details]{Novick72, Weisskopf72}. Since that first successful measurement, several high-energy missions were planned to include X-ray polarimeters (e.g. \emph{Einstein} Observatory, Spectrum-X). However, none of these polarimeters were launched. Developing a polarimeter capable of measuring linear polarization of astronomical sources in the range $0.1 - 50$ keV has many challenges and two different types of polarimeters have been proposed: scattering polarimeters and photoelectron polarimeters \citep[see][for a detailed description]{Weisskopf06}. Each type of detector has its own limitations, although photoelectric X-ray polarimeters with gas detectors have recently reached a mature level of development and variations of this model have been proposed for a number of future missions \citep[for e.g. see][]{Costa01, Bellazzini07, Gunji04, Jahoda07, Arimoto08}. To date, however, very little theoretical work has been done on this subject. Indeed, the discriminating polarization signatures of the most powerful cosmic X-ray sources has not yet been determined.

We have developed a Monte Carlo algorithm to model polarized photon transport resulting from Compton scattering which we have applied to a magnetic cataclysmic variable (mCV) accretion column \citep*{McNamara08a}. We demonstrated that the X-ray polarization levels are significantly higher in a shock heated accretion column which is stratified in density and temperature than in a uniform column \citep{Matt04}. We also demonstrated that the degree of polarization depends on the emission sites in the source region. We now extend our study to model X-ray polarization in relativistic jets. 
Although our focus is on jets in AGN, the results obtained here are also applicable to relativistic jets in galactic X-ray binaries \citep[e.g.][]{Fender06} and ultra-luminous X-ray sources \citep[e.g.][]{Freeland06}. 

Jets in radio-loud active galatic nuclei (AGN) have been studied for the last 25 years with radio and optical observations \citep[e.g.][]{Bregman90, Tavecchio07}. It is only since the launch of the \emph{Chandra} X-ray observatory, with sub-arcsecond angular resolution, that jets have become an important topic in X-ray astronomy. \emph{Chandra} has provided imaging spectroscopy studies of extended jets, revealing that jet X-ray emission is much more complex than previously thought \citep[see][for X-ray jet surveys]{Marshall05,Sambruna04}. In particular, the origin of the X-ray emission is not clear from the available data and considerable effort has gone into explaining the featureless power-law spectrum seen in all these sources. X-ray emission in relativistic jets in AGN may arise from a number of different processes. Polarization studies in the radio and optical bands, indicate that this emission mainly originates from synchrotron radiation \citep{Jorstad07}. Thus, synchrotron and synchrotron self-Compton (SSC) emission are obvious candidates for the X-ray continuum emission \citep{Maraschi92}. However, jet X-ray emission may also originate from external Comptonization (EC) of disk blackbody radiation \citep{Dermer93, Wagner95} or of the cosmic microwave background (CMB) \citep{Tavecchio00, Celotti01}. X-ray polarization measurements may be able to provide an independent diagnostic for discriminating between these competing emission mechanisms. 

While the polarization properties of synchrotron emission are well known, only a few approximate analytical predictions have been made for SSC polarization \citep{Bjornsson82b, Begelman87, Celotti94} or for external Comptonized emission \citep{Poutanen94}. In this paper, we calculate the X-ray polarization arising from photons scattered by energetic electrons in jets at relativistic bulk speeds. We consider Compton scattering of thermal photons emitted from an underlying accretion disk as well as scattering of the intrinsically polarized synchrotron photons emitted within the jet. We also examine the effects of Compton scattering of CMB photons within the jet. The paper is outlined as follows: in Section \ref{theory} we describe the jet model, outline the theory of polarization due to Compton scattering and describe our computational algorithm. In Section \ref{results} we present our Monte Carlo modelling results for EC polarization and SSC polarization. We summarise our results in Section \ref{conclusion}.

\section{Theory}
\label{theory}
\subsection{The Jet Model}

\begin{figure}
	
	\includegraphics[width=9.0truecm]{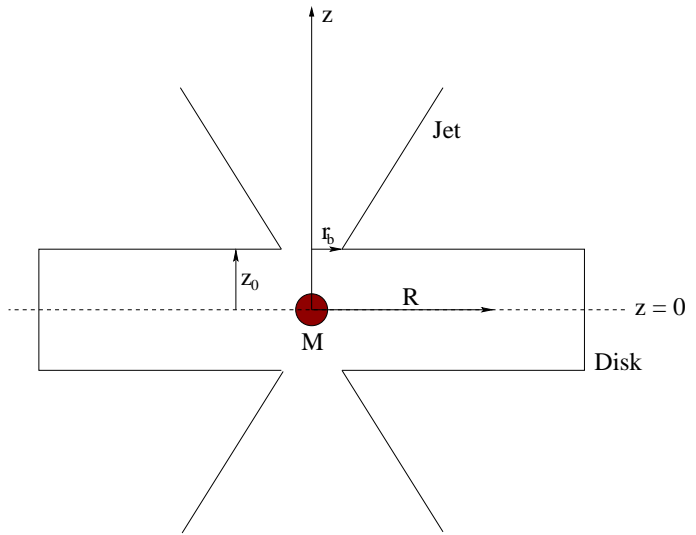}
	\caption{Schematic geometry of a relativistic jet and black hole accretion disk model: $M$ is the black hole mass, $z_{\rm 0}$ is the height above the disk midplane ($z = 0$) where the jet is launched and $r_{\rm b}$ is the radius at the base of the jet.}
	\label{altgeo}
	
\end{figure}	

We consider a relativistic jet with Lorentz factor $1 < \Gamma_{\rm j} < 10$ for central object masses $M = 10^6 -10^8 \Msun$ with a mass accretion rate $\dot M_{\rm a}$. The jet is modelled with a conical shape as shown in Fig.~\ref{altgeo} and is launched at a height $z_{\rm 0}$ above the disk midplane. The jet electrons have a non-thermal powerlaw energy distribution, 
\begin{equation}
	N_{\rm e} (\gamma) = K\gamma^{-p}
\label{powerlaw}	
\end {equation}
where $N_{\rm e}$ is the electron number density, $\gamma$ is the electron Lorentz factor, $p$ is the particle spectral index and $K$ is a normalization factor obtained from $N_{\rm e} = \int_{\gamma_{\rm min}}^{\gamma_{\rm max}} N_{\rm e} (\gamma) \, d\gamma$. The electron number density at $z_{\rm 0}$ can be calculated from energy conservation assuming that the bulk kinetic energy dominates \citep[e.g.][]{Celotti98, Freeland06},
\begin{equation}
	N_{\rm 0} = N_{\rm e}(z_{\rm 0}) \approx \frac{P_{\rm j}}{\Gamma_{\rm j}\beta_{\rm j} c(\Gamma_{\rm j} -1)m_{\rm p}c^2\pi r_{\rm b}^2} \qquad .
\end{equation}	
Here, $P_{\rm j}$ is the jet power and $r_{\rm b}$ is the radius of the base of the jet.   
We let the electron number density in the jet fall off according to mass continuity, 
\begin{equation}
	N_{\rm e}(z) = N_{\rm 0}\left(\frac{z}{z_{\rm 0}}\right)^{-2} \quad .
\label{numberdens}	 
\end{equation}

\subsection{Accretion Disk and CMB Model}

We consider the accretion disk model of \cite{SS73}. A blackbody temperature is determined for each disk annulus $R + \delta R$, from $\sigma T^4(R) = F(R)$ where $\sigma$ is the Stefan-Boltzmann constant and
\begin{equation}
	F(R) = \frac{3GM\dot M_{\rm a}}{8\pi R^3} \left[1 - \left(\frac{R}{R_{\rm i}}\right)^{-1/2}\right]  \qquad .
\label{Flux}	
\end{equation}	
Here $R_{\rm i}$ is the innermost stable orbit of the disk. Photons emitted by the disk may enter the jet and scatter with the jet electrons (see Fig.~\ref{emissiontype}). 

We also investigate the scattering of CMB blackbody photons in the jet. The seed CMB photons are selected from a single-temperature blackbody. In this work we consider jets at redshift of $z=2$. We note that the results can be generalized to higher or lower red-shifts simply by adjusting the temperatures of the injected blackbody photons.

\subsection{Polarized Compton Scattering}

The differential Klein-Nishina cross-section for scattering of polarized electromagnetic waves, with polarization vector $\mathbf{e}$, off electrons in the electron rest frame is \citep[see][for example]{Heitler36, Jauch80}, 
\begin{equation}
  \left(\frac{d\sigma}{d\Omega}\right)_{\rm e} = \frac{1}{2}r_{0}^2\eta^2_{\rm e}  \left[ \eta_{\rm e} + \eta_{\rm e}^{-1} -2 \sin^2 \theta_{\rm e} \cos^2 \phi_{\rm e} \right] \qquad .
  \label{P}
 \end{equation}
Here, $r_{0}$ is the classical electron radius and
\begin{equation}
 \eta_{\rm e} \equiv \left(\frac{\nu'}{\nu}\right)_{\rm e}  = \left[1 + \frac{h\nu_{\rm e}}{m_{\rm e} c^2}(1-\cos\theta_{\rm e}) \right]^{-1} \qquad .
\end{equation}
The subscript 'e' denotes quantities calculated in the electron rest frame and primed quantities denote values after scattering. Here, $\nu$ is the frequency of the incident photon, $m_{\rm e}$ is the mass of an electron, $\theta$ is the scattering angle (angle between the incident and scattered photon wavevector) and $\phi$ is the azimuthal angle.

The polarization vector $\mathbf{e}\,'$ for the polarized fraction $P$ of scattered photons is perpendicular to the scattering plane and is defined by \citep{Angel69},
\begin{equation}
  \mathbf{e}\,' = \frac{1}{|\mathbf{e}'|} (\mathbf{e} \times \hat{\mathbf{\Omega}}') \times \hat{\mathbf{\Omega}}' 
  \label{Angeleqn}
\end{equation}
where $\hat{\mathbf{\Omega}}'$ is the propagation directional unit vector of the scattered photon. For the remaining $1-P$ fraction of photons $\mathbf{e}\,'$ is randomly distributed in the plane perpendicular to $\hat{\mathbf{\Omega}}'$. In the case where the radiation is intially unpolarized, the degree of linear polarization induced by Compton scattering is given by \citep[see e.g.][]{Dolan67, Matt96}, 
\begin{equation} 
  P = \sin^2\theta_{\rm e} \left( \eta_{\rm e} + \eta^{-1}_{\rm e} -1 + \cos^2\theta_{\rm e}\right)^{-1}   \quad .
\label{P}  
\end{equation}
For polarized incident emission, the degree of linear polarization induced is given by, 
\begin{equation}
  P = 2\frac{1-\sin^2\theta_{\rm e}\cos^2\phi_{\rm e}}{\eta_{\rm e} + \eta_{\rm e}^{-1} - 2\sin^2\theta_{\rm e}\cos^2\phi_{\rm e}} \quad .
\end{equation}
Further details can be found in \cite{McNamara08a}.

\subsection{Computational Algorithm}

\begin{figure*}
	(a) 
	\includegraphics[width=7.0truecm]{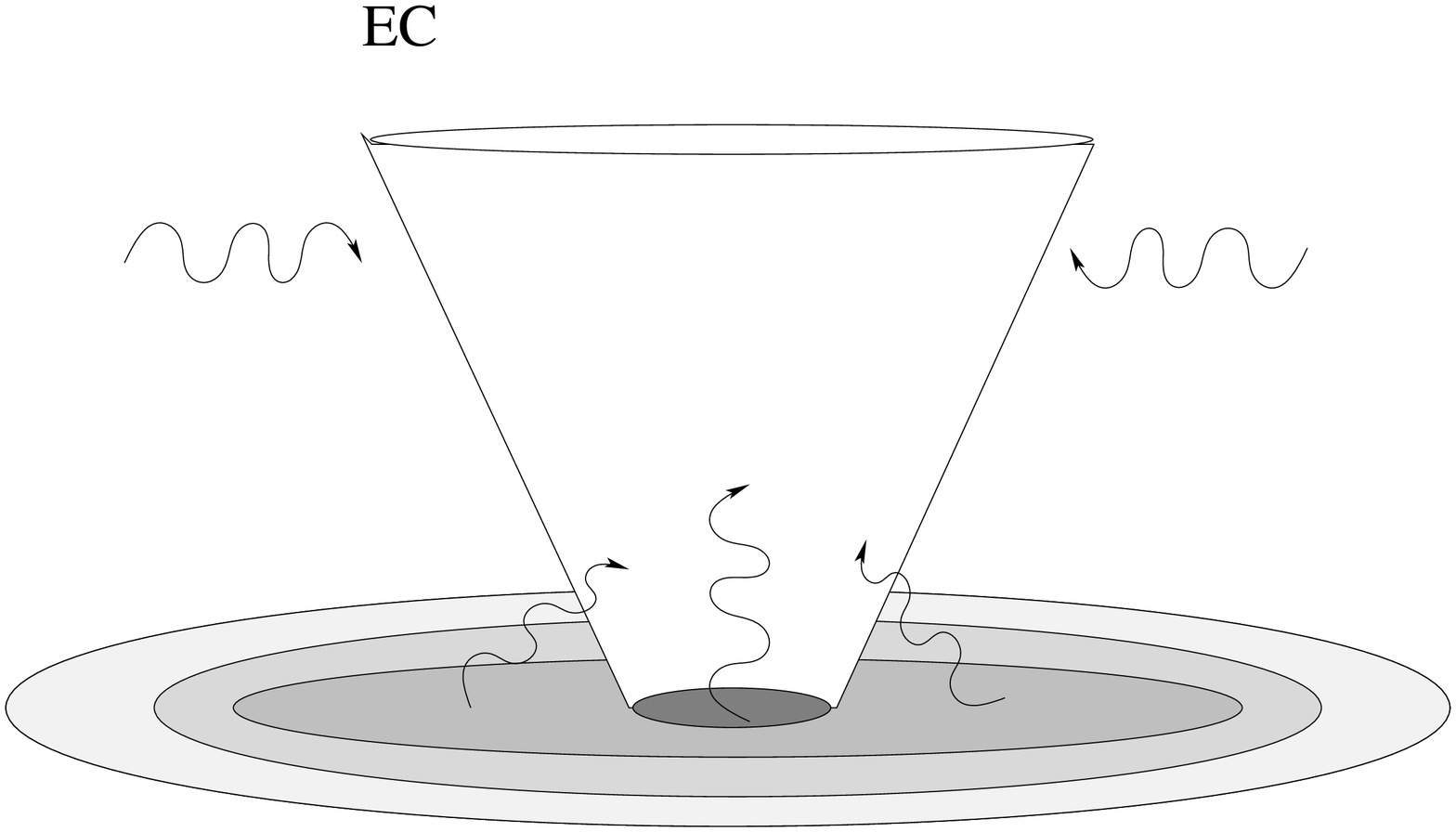}
	(b) 
	\includegraphics[width=7.0truecm]{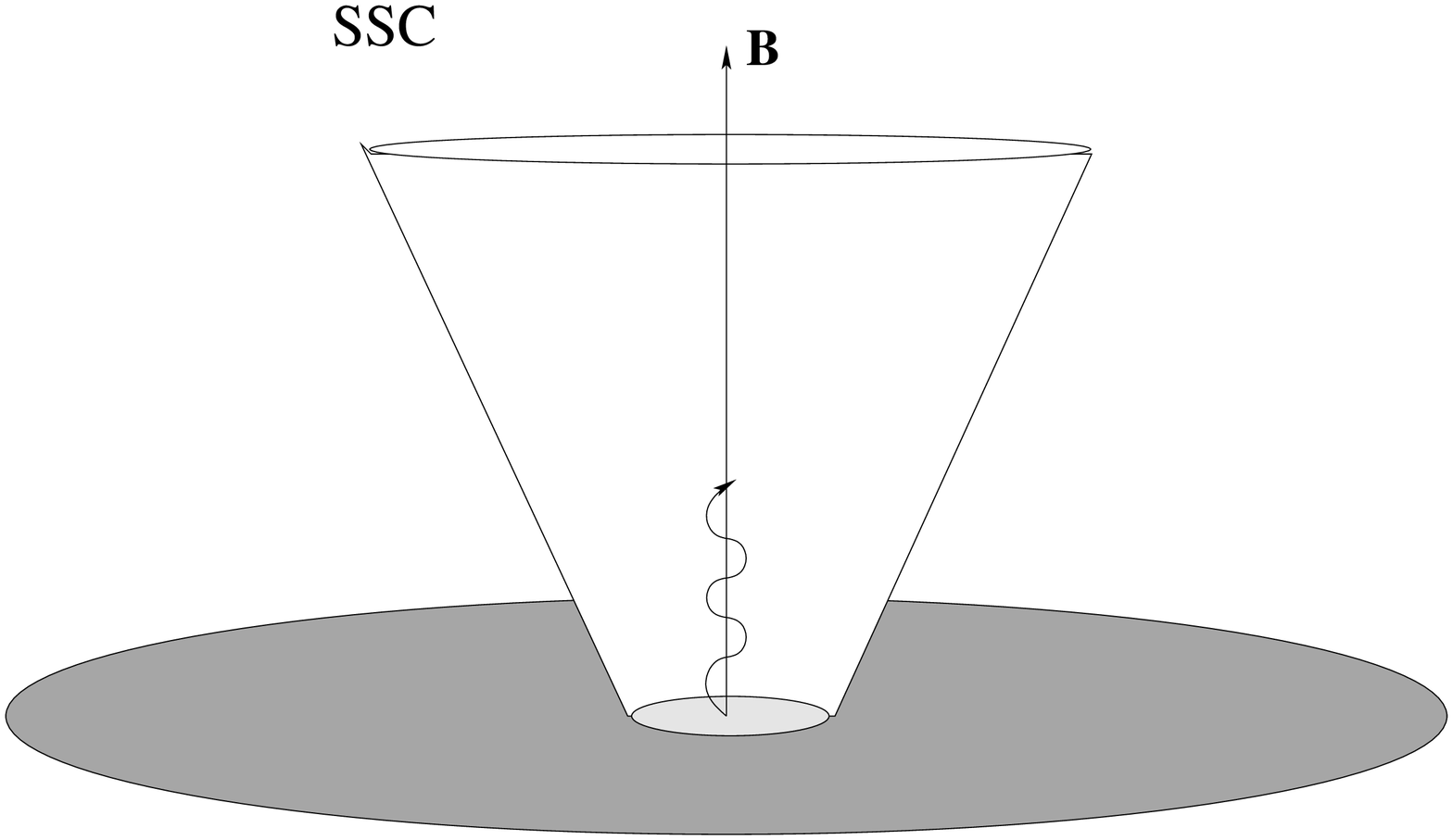}
	\caption{An illustration of the possible different types of X-ray emission mechanisms in a jet. (a) External Comptonization (EC) of seed photons from an accretion disk and/or from the CMB; (b) Synchrotron Self-Comptonization (SSC) photons emitted by the jet plasma.  }
	\label{emissiontype}
	
\end{figure*}

Three different reference frames are used in the computation: the plasma rest frame (PF) which is the frame co-moving with the emitting jet plasma; the electron rest frame (EF) and the observer rest frame (OF). In the EF the $z$-axis is defined by the direction of the electron propagation $\hat{\bbeta}_{\rm e}$, while in the PF the $z$-axis is along the magnetic field of the jet. In the case where the PF and the EF are not in a standard configuration (i.e. the EF does not move along the $z$-direction of the PF and the corresponding axes of the two frames are not parallel) a new rotated frame in standard configuration is first defined in the PF and then transformations between frames are performed. 

We model the scattering of three different types of photons in our analysis: unpolarized blackbody disk photons, unpolarized blackbody CMB photons and intrinsically polarized synchrotron photons. 

\subsubsection{Blackbody Disk and CMB Photons}

The disk photons are emitted in the OF from a multicolour disk model. We assume that the photons are intially unpolarized. The probability density of a disk seed photon can be written as,
\begin{equation}
p(R) = 4\pi R dR \frac{F(R)}{L}
\end{equation} 
where $F(R)$ is the flux density at disk radius $R$ (see equation \ref{Flux}) and $L$ is the total luminosity of the disk. We sample $p(R)$ by tabulation and determine the blackbody temperature at $R$. The photon energy is then sampled from the Plank function $B_{\nu}(R)$ as described in \citet{P83}. Some of the disk photons may impact the jet and undergo scattering, these photons are transported through the jet until they escape. 

CMB photons may also undergo scattering within the jet (see Fig.~\ref{emissiontype} (a)). We sample the CMB photons from a blackbody distribution with a temperature $T_{\rm bb} = 2.8 \, (1+z)$~K in the OF at a redshift $z$. The initially unpolarized photons are emitted isotropically around the jet. If a photon enters the jet it may scatter with the jet electrons and become polarized.   

\subsubsection{Synchrotron photons}

Synchrotron radiation is emitted by the jet electrons with electron Lorentz factors $\gamma_{min} \leq \gamma \leq \gamma_{\rm max}$. We emit the photons isotropically in the EF with an initial energy chosen from a powerlaw distribution. Synchrotron radiation is polarized and depends on the magnetic field $\mathbf{B}$ of the jet, thus the initial polarization vector $\mathbf{e}$ is chosen in the PF so that it is perpendicular to both $\bmath{B}$ and $\bmath{\Omega}$. We Lorentz transform the synchrotron photons from the EF to the PF using, 
\begin{equation}
	\bmath{\Omega}_{\rm pf} = \delta_{\rm e}[\bmath{\Omega}_{\rm ef} - \gamma_{\rm e}\bmath{\beta}_{\rm e} - (\gamma_{\rm e} - 1)(\hat{\bmath{\Omega}}_{\rm ef} \cdot \hat{\bmath{\beta}}_{\rm e})\hat{\bmath{\beta}}_{\rm e}] 
	\label{Lorentzprop}
\end{equation}	
and
\begin{equation}
	E_{\rm pf} = \frac{E_{\rm ef}}{\delta_{\rm e}}
	\label{LorentzEnergy}
\end{equation} 
where the Doppler factor $\delta_{\rm e} = \{ \gamma_{\rm e}[1 - \beta_{\rm e}(\hat{\bmath{\Omega}}\cdot\hat{\bmath{\beta}}_{\rm e})]\}^{-1}$ and 'pf' and 'ef' denote quantities in the plasma and electron rest frame, respectively.  

\subsubsection{Transport and scattering of photons}

Once the incident photon energy, direction and polarization have been selected, the photon is transported through the jet and the distance the photon has travelled is determined. If the photon encounters an electron before leaving the jet, we determine whether scattering occurs using a rejection algorithm \citep{Cullen01a, Cullen01b}. If the scattering event is rejected, the photon continues in the same direction without scattering. If the scattering event is accepted, the photon energy and $\mathbf{\Omega}$ is Lorentz transformed into the electron rest frame.

The scattered photon quantities are selected by determining the scattered azimuthal angle $\phi_{\rm e}'$ and the cosine of the angle between the scattered photon and electron $\mu_{\rm e}$ \citep[for a detailed description of the Compton scattering algorithm see,][]{McNamara08a, McNamara08b, Kuncic05}. Next the scattered photon direction, scattering angle $\theta_{\rm e}$ and photon energy are selected. Another rejection algorithm determines whether the scattered quantities should be accepted or chosen again. If the scattered quantities satisfy the selection criteria, we determine whether the scattered photon has been polarized and calculate the scattered polarization vector $\bmath{e}'$ using (\ref{Angeleqn}). If the photon is unpolarized after scattering, $\bmath{e}'$ is chosen randomly with the condition $\bmath{e}'\cdot\bmath{\Omega}' = 0$.

Finally, the photon properties are transformed into the observer's frame using, 
\begin{equation}
	\bmath{\Omega}_{\rm of} = \delta_{\rm j}[\bmath{\Omega}_{\rm pf} + \Gamma_{\rm j}\bmath{\beta}_{\rm j} + (\Gamma_{\rm j} - 1)(\hat{\bmath{\Omega}}_{\rm pf} \cdot \hat{\bmath{\beta}}_{\rm j})\hat{\bmath{\beta}}_{\rm j}] 
	\label{Lorentzpropj}
\end{equation}	
and
\begin{equation}
	E_{\rm of} = \frac{E_{\rm pf}}{\delta_{\rm j}}
	\label{LorentzEnergyj}
\end{equation}   
where $\delta_{\rm j} = \{ \Gamma_{\rm j}[1 - \beta_{\rm j}(\hat{\bmath{\Omega}}\cdot\hat{\bmath{\beta}}_{\rm j})]\}^{-1}$ is a Doppler factor and 'of' denotes quantities in the observer frame. $\bmath{\beta}_{\rm j}$ is the relative velocity of the jet w.r.t. the observer. 

When Lorentz transforming the polarization vectors we work with the 4-vector $\epsilon$ in the temporal gauge where $\epsilon^0 = 0$. While transforming the polarization 4-vector to a different frame, we use the following prescription; make a Lorentz transformation to find $\epsilon$ in the new frame and then make a gauge transformation to satisfy the gauge conditions $\epsilon = [0, \bmath{e}]$ and $\bmath{e} \cdot \bmath{\Omega} = 0$. 

\section{Results and discussion}
\label{results}

We model a relativistic jet as described in Section~\ref{theory}. We consider a system with central object mass $M = 10^8 \, \Msun$, jet Lorentz factor $\Gamma_{\rm j} = 5$, jet opening angle $\omega\sim 1/\Gamma_{\rm j} \sim 11\degree$  and a mass accretion rate $\dot M = 0.1 \, \Msun \, \rm yr^{-1}$. For these quantities, the optical depth across the base of the jet is $\tau_{\rm 0} \approx 4 \times 10^{-4}$. We assume that 50\% of the accretion power is chanelled to the jet. The jet is observed from an inclination angle $i$ measured with respect to its axis and photons are binned for a small range of inclination angles $\delta i \approx 5\degree$. We consider jet electron Lorentz factors $\gamma_{\rm min} \le \gamma \le \gamma_{\rm max}$, with $\gamma_{\rm min} = 1$ and $\gamma_{\rm max} = 10^5$ and a particle spectral index $p = 3$. The jet geometry is conical with a base radius $r_{\rm b} = 10$~$\rm r_{\rm g}$ and total jet length $z \approx 500$ $\rm r_{\rm g}$, where $r_{\rm g} = GM/c^2$. The jet is launched at a height $z_{\rm 0} = 0.5 \, r_{\rm g}$ (see Fig.~\ref{altgeo}). 

We investigate the X-ray polarization and spectra of three different emission models: EC of disk and CMB photons as well as SSC photons. For the disk Comptonization model, we use a multi-colour disk spectrum \citep{SS73}. For the CMB Comptonization model, we assume the jet is at a redshift of $z=2$. 

An alternative EC model which may produce X-rays through inverse Compton emission in a jet, is Comptonization of seed photons from the broad-line region (BLR) or from a dusty torus \citep[see][]{Sikora94}. It is likely that the X-ray emission in low-luminosity sources is dominated by SSC, while emission from high-luminosity sources is largely due to a combination of synchrotron and EC, including EC of BLR and torus photons \citep[e.g.][]{Harris06}. The X-ray polarization of jet Comptonized BLR or torus photons is likely to be similar to that of scattered CMB photons since they have a similar isotropic distribution. However, as these regions are much further away from the primary X-ray source region, their contribution to the overall X-ray polarization may be less important than that of SSC or other EC mechanisms (disk, CMB). Futhermore, additional scattering and source regions, such as the BLR and torus, are expected to produce more complicated polarization signatures across several wavebands. While this could be important for constraining the size and geometry of these regions in some sources, such a study is beyond the scope of this paper. 

We adopt the general convention used for the polarization degree $P$, assuming it is positive when the polarization is perpendicular to the projection of the jet axis onto the plane of the sky and negative when it is parallel. We use $10^9$ photons in each simulation to determine $P$. 

\subsection{Disk Comptonization}

\begin{figure}
	
	\includegraphics[width=9.0truecm]{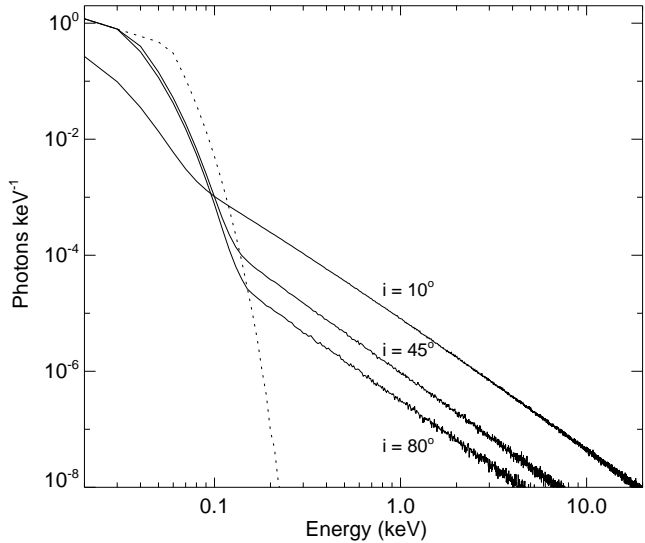}
	\caption{Simulated spectra of disk blackbody photons scattered in a relativistic jet for three different inclination angles $i$. The dotted line shows the input seed spectrum from a multi-colour disk model.}
	\label{BBspectrum}
	
\end{figure}	

\begin{table}
\caption{The polarization degree $P$ for disk photons scattered to X-ray energies in a relativistic jet seen at an inclination angle $i$ and the average number of scatterings each photon undergoes before escaping the jet.} 

\begin{tabular}{ccc}
\hline
$i$   & $P \,(\%)$            & Average number of      \\ 
      & (E = $1 -10$ keV )    & scatterings per photon \\ 
\hline
$10\degree$  & $3.2$   & $3.0$\\
$45\degree$  & $14.0$  & $2.8$\\
$80\degree$  & $20.6$  & $2.8$\\

\hline
\hline
\end{tabular}
\label{BBpolarization}
\end{table}

Blackbody photons emitted from the accretion disk may enter the jet and scatter. A combination of factors determines the overall polarization of the resulting X-ray photons. In particular, since the disk photons are initially unpolarized, the number of scatterings and the angle a photon scatters into will be important in contributing to the net X-ray polarization. Fig.~\ref{BBspectrum} shows the scattered spectrum for three different inclination angles: $i = 10\degree$, $i = 45\degree$ and $i = 80\degree$ (solid curves) and the input seed spectrum (dotted curve).  Table~\ref{BBpolarization} tabulates the polarization degree of disk photons with energies between $1-10$ keV seen at a particular $i$. Also tabulated are the average number of scatterings each photon undergoes before escaping the jet. 

Since the seed photons have small energies $\le 0.1$ keV, all photons observed with energies $1 - 10$ keV must have undergone at least one scattering. The optical depth across the jet is relatively small, $\tau_{\rm 0} \sim 4 \times 10^{-4}$ at the jet base, so photons traversing across the base are practically unscattered. However, the majority of the disk seed photons enter the jet from directions approximately parallel to the jet axis. The effective scattering probability is substantially higher than that indicated by the scattering optical depths across the jet base. 
Photons propagating along the jet axis is more likely to be scattered and become polarized. Moreover, the photons escaping at locations high above the jet base should experienced more scatterings. Photons emerging at large $i$ tend to have large deflection angles in their very last scatterings due to the large aspect ratio of the jet and low scattering optical depth across the jet. This together with the fact that the injected photons are unpolarized leads to an increase in the fractional polarization with the viewing inclination angles $i$.    

Another important effect evident in Fig.~\ref{BBspectrum} is bulk Doppler boosting. The disk photons entering the jet are preferentially scattered in the forward direction of the bulk jet plasma, as viewed in the observer's rest frame. The observer sees more energetic photons at small inclinations (see the $i = 10\degree$ curve in Fig~\ref{BBspectrum}). The observed spectrum also contains the disk photons that leave the disk at an inclination $i$ but do not enter and scatter within the jet. For $i=10\degree$ a larger fraction of seed photons are upscattered because they are more likely to enter the jet than disk photons emitted at larger angles. 

\subsection{CMB Comptonization}

\begin{figure}
	
	\includegraphics[width=9.0truecm]{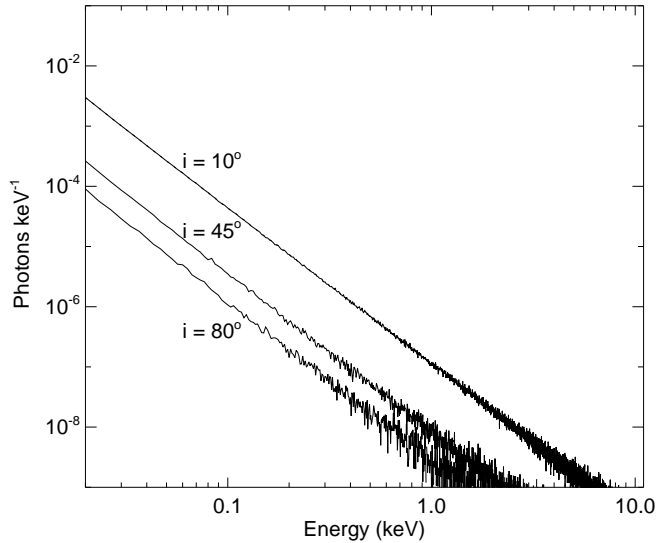}
	\caption{Simulated spectra of CMB photons scattered within a relativistic jet for three different inclination angles $i$ for a source at $z =2$. The seed photons are selected from a single temperature blackbody distribution.}
	\label{CMBspectrum}
	
\end{figure}
	
\begin{table}
\caption{The polarization degree $P$ for CMB photons scattered to X-ray energies in a relativistic jet seen at an inclination angle $i$ and the average number of scattering each photon undergoes before escaping the jet.} 
\begin{tabular}{ccc}
\hline
$i$   & $P\,(\%)$          & Average number of       \\ 
      & (E = $1 -10$ keV ) & scatterings per photon  \\ 
\hline
$10\degree$  & $4.2$    &$3.2$   \\
$45\degree$  & $16.5$   &$2.6$   \\
$80\degree$  & $23.9$   &$3.2$   \\
\hline
\hline
\end{tabular}
\label{CMBpolarization}
\end{table}

The CMB photons are selected from a single-temperature blackbody distribution and emitted isotropically in the area surrounding the jet. Since external Comptonization of CMB photons may only contribute to the observed X-ray spectrum for high redshift sources (because of the $(1 + z)^4$ dependence of the energy density), we consider a source at $z=2$ \citep{Schwartz02}. The jet then sees a CMB temperature of $8.346$ K ($7.2 \times 10^{-4}$ eV). Fig.~\ref{CMBspectrum} shows the simulated X-ray spectra for the upscattered CMB photons for three different jet inclination angles: $i = 10\degree$, $45\degree$ and $80\degree$. Table~\ref{CMBpolarization} shows the polarization degree $P$ for the CMB photons scattered to energies $1-10$ keV, for each $i$. Also tabulated is the average number of scatterings each photon undergoes. As the CMB seed photons have average energies lower than the disk seed photons, the average number of scatterings needed to reach X-ray energies is expected to be higher.  

As is evident in Table~\ref{CMBpolarization}, the X-ray polarization of the scattered CMB photons follows the same trend with viewing inclination angle as for the case of disk Comptonization. However, the predicted $P$ at each $i$ are slightly different. We attribute this to the difference in the angular distribution of photon injection and the polarization condition in the last scattering event. The disk photons are unpolarized and emitted near the base of the jet and hence the disk seed photons participate in the scattering process have a narrow range of initial pitch angles. Given the relatively low scattering optical depth in the jet, photons that have a single scattering dominate and hence photons emerging at small $i$ are mostly restricted to small-angle scatterings (analogous to total internal reflection). The CMB photons are unpolarized, similar to the disk photons. However, the angular distribution of the CMB seed photons involved in the scattering is isotropically around the whole length of the jet, which is contrary to the more restrictive angular distribution of the disk seed photons. For CMB scattering, photons emerging at low $i$ consist of two main types: single scattered photons with large deflection angles, and multiple scattered photons with small deflection angles in the last scattering. For the case of the single scattered photons, the injected photons are unpolarized, and the polarization is caused by large-angle scatterings. In the case of multiple scattered photons, the pre-scattered photons, which propagate approximately along the jet, are already polarized because of previous scatterings. Thus, the polarization is slightly higher for scattering of CMB photons than disk photons. Scattered CMB photons emerging at large $i$ mostly undergo many scattering with a large deflection angle in the last scattering. As the pre-scattered photons in the last scattering are polarized, it is not surprising that the resulting polarization is higher than those of the disk photon scattering, whereas the pre-scattered photons in the last scattering is less polarized.  
   
The differences in the X-ray polarizations for the EC disk and CMB models are only 1\% at $i = 10\degree$ and $<3.5$\% at $i = 80\degree$. Such a small difference makes it difficult to discriminate between these two models from X-ray polarization observations alone.     

\subsection{Synchrotron-Self Comptonization} 

We expect the SSC polarization signature to differ significantly from the EC case, since synchrotron photons emitted by jet electrons are intrinsically polarized. In the EC case, Compton scattering polarizes the seed photons, whereas in the case of SSC, scattering may depolarize the photons. 

The polarization vector $\mathbf{e}$ of the incident synchrotron photons is chosen so that it is perpendicular to both the magnetic field of the jet and the photon propagation vector $\mathbf{\Omega}$ in the PF. We consider a simple model, where the magnetic field is directed along the jet axis and we assume that the jet electrons spiral around the $z$-axis in the PF. The intrinsic polarization in this case will be perpendicular to the projection of the jet axis in the sky. Realistically, the magnetic field may have a more complicated configuration which could produce different polarization signatures. Here we demonstrate that the polarization of the SSC photons may be distinguishable from that of the EC emission even though the actual polarization signatures may differ for more complicated geometries and magnetic fields. 

We investigate the effect of different emission sites for the primary synchrotron photons. We first inject the seed photons uniformly throughout the jet and then from a particular dimensionless height $\zeta$, where $\zeta = 1$ and $\zeta = 0$ corresponds to the top and base of the jet, respectively. Uniform emission of synchrotron seed photons in the jet is consistent with some observations of radio jets \citep[e.g. M87,][]{Kovalev07}. It may be attributed to continuous acceleration of electrons throughout the jet as a result of stochastic reconnection events in a random magnetic field component \citep[e.g.][]{Blandford87}. On the other hand, emission of synchrotron photons at localized sites within the jet could occur as a result of discrete shock events. There is some evidence for this from observations of bright spots in radio jets \citep[e.g.][]{Kataoka05}. 

\begin{figure}
	
	\includegraphics[width=9.0truecm]{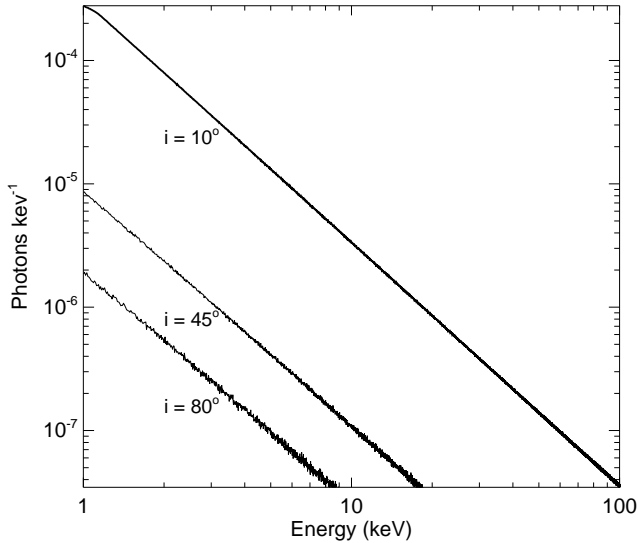}
	\caption{Simulated spectra of SSC photons for three different observer viewing angles $i = 10\degree$, $45\degree$ and $80\degree$.}
	\label{SSCspectrum}
	
\end{figure}

Fig.~\ref{SSCspectrum} shows the simulated spectra of the total (seed synchrotron plus SSC) photons emerging at three different observer viewing angles $i$. The seed synchrotron photons are emitted uniformly throughout the jet in this case. Boosting occurs along the jet axis, hence most photons are observed at small inclinations ($i = 10\degree$ case in Fig.~\ref{SSCspectrum}).        

\begin{figure}
	
	\includegraphics[width=9.0truecm]{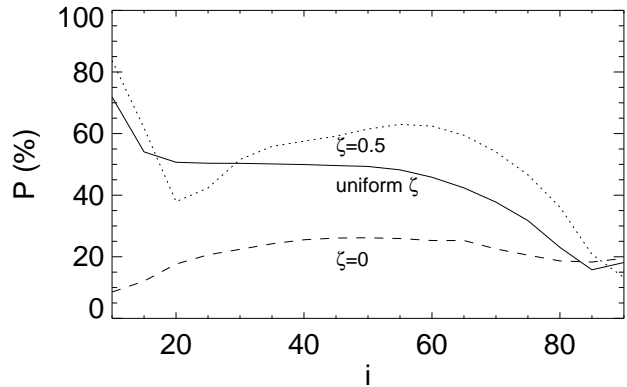}
	\caption{Polarization degree $P$ of SSC photons with energies between $1-10$~keV plotted as a function of the inclination angle $i$. The solid line is for the case where the seed photons are emitted uniformly throughout the jet (uniform $\zeta$). The dashed and dotted lines are for the cases where the seed photons are emitted at the jet base ($\zeta =0$) and in the middle of the jet ($\zeta = 0.5$).}
	\label{SSCpol}
	
\end{figure}

\begin{figure}
	
	\includegraphics[width=9.0truecm]{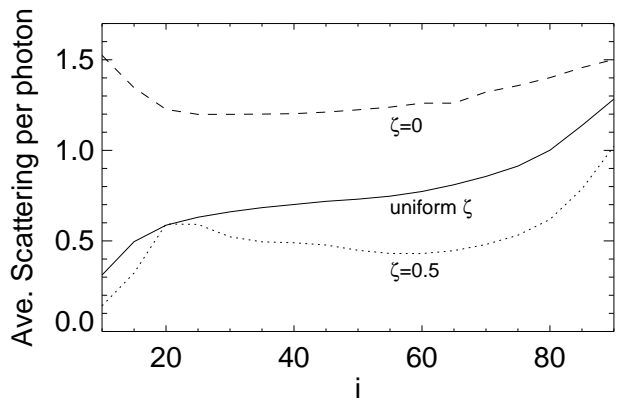}
	\caption{The average number of scatterings each SSC X-ray photon undergoes before escaping the jet as a function of inclination angle $i$. The SSC photons (see Fig.~\ref{SSCspectrum}) are emitted either uniformly throughout the jet (uniform $\zeta$), at the base of the jet ($\zeta = 0$) or in the middle of the jet ($\zeta = 0.5$).}
	\label{SSCScatnum}
	
\end{figure}

Fig.~\ref{SSCpol} shows the polarization degree $P$ of photons with energies between $1-10$~keV emerging from a jet at different inclination angles $i$ and for different synchrotron injection sites $\zeta$. Fig.~\ref{SSCScatnum} shows the corresponding average number of scatterings each photon undergoes before escaping the jet at $i$. There is a clear correlation between number of scatterings and depolarization of the seed photons. As the seed synchrotron photons are emitted into a narrow emission cone about each electron's instantaneous propagation direction along the $z$-axis (in the PF), photons which are observed to emerge from the jet at large $i$ must undergo a larger number of scatterings (see Fig~\ref{SSCScatnum}). Photons emerging from small $i$ may also experience a large number of scatterings when emitted from close to the jet base because these photons propagate along the entire jet axis and thus see a large optical depth.

Figs.~\ref{SSCpol} and \ref{SSCScatnum} show that the polarization of SSC jet photons is sensitive to the emission site of the seed photons. We investigate three different cases: photons emitted from localized regions at the jet base ($\zeta = 0$) and at the middle of the jet ($\zeta = 0.5$) and photons emitted uniformly throughout the jet. The overall polarization of photons emitted at the jet base is significantly lower than in the other two cases, because the optical depth is largest at the jet base, where $N_{\rm e}$ is highest, and thus seed photons emitted from $\zeta = 0$ have a higher probability of scattering and depolarizing. Photons emerging from the jet at $i = 10\degree$ have the lowest polarization since they travel across the longest geometric path, along the entire jet length. Photons emitted from source sites higher up in the jet experience fewer scatterings on average and consequently, the X-ray polarization is higher than that for synchrotron source sites closer to the jet base which suffer stronger SSC depolarization (compare the $\zeta = 0.5$ and $\zeta = 0$ curves in Figs.~\ref{SSCpol} and \ref{SSCScatnum}). This suggests that X-ray polarization offer clues to locating particle acceleration sites in jets. In the $\zeta = 0.5$ case, the sharp decrease in $P$ for $i \approx 20\degree$ arises from the conical jet geometry we have considered. For a given seed photon localized injection site $\zeta$, and for escaping angles larger than the jet opening angle, there is a propagation direction for which a photon sees a maximum optical depth (optimized by the combination of $N_{\rm e}$ and the geometric path length). Thus, there is a higher probability of scattering along this path compared to the other paths of photons injected from the same site. For $\zeta = 0.5$, the maximum optical depth occurs for photons emerging at $i\sim 10\degree$. For injection sites higher up in the jet, where $N_{\rm e}$ is lower, the maximum optical depth will occur on a longer geometric path and will thus correspond to a different photon escape angle $i$. 

For the case where photons are emitted uniformly throughout the jet length (solid curves in Fig.~\ref{SSCpol} and \ref{SSCScatnum}), the X-ray polarization is approximately a combination of that for different localized injection sites. 

\section{Conclusion}
\label{conclusion}

We have investigated the properties of Compton polarized X-rays from a relativistic jet using Monte Carlo simulations. We consider three different emission mechanisms proposed for the observed X-ray emission in AGN: external Comptonization of blackbody accretion disk radiation and CMB emission as well as SSC emission. 

In the case of the initially unpolarized EC emission we find that the deflection angle in the last scattering is the most important factor in determining the X-ray polarization for a particular viewing angle $i$. Photons that undergo larger angle scatterings result in a higher polarization than those undergoing smaller angle scatterings. In the case of disk photons, the photons enter the jet with initial propagation directions mostly along the jet axis and thus, we find that photons scattering into large viewing angles have higher polarization, $P \approx 20 \%$. CMB photons enter the jet quasi-isotropically and uniformly along its length. In this case, because of the larger distribution of injection angles, some photons seen at small $i$ will have undergone large angle scatterings and as a result $P$ is slightly higher than in the disk EC case for small $i$. Both EC models show increasing $P$ with $i$. The polarization degrees have similar magnitudes and our results suggest that it would be difficult to discriminate between the the two EC models with X-ray polarization studies alone.

In the case of SSC, we find that depolarization of synchrotron photons is sensitive to the location of the seed photon emission site. Synchrotron photons emitted near the jet base generally suffer stronger SSC depolarization than those emitted higher up or uniformly throughout the jet. This is because photons emitted near the jet base generally see a larger optical depth and hence, undergo more scatterings. This is especially the case for photons escaping at small $i$; in this case, the X-ray polarization is expected to be $P \leq 20\%$ if the seed synchrotron photons orginate near the jet base, or $P \geq 50\%$ if they are emitted uniformly throughout the jet. This implies that X-ray polarimetry can discern the scattering site and hence constrain models for particle accleration in relativistic jets. 

In summary, polarimetric observations would be able to discrimate between synchrotron/SSC and EC X-ray emission (and also between synchrotron and SSC X-ray emission). However, it would be difficult to identify the type of EC emission (disk or CMB), from X-ray polarimetric observations alone.  

\section*{Acknowledgments}
ALM thanks a University of Sydney Denison Scholarship. 
ZK acknowledges support from the Australian Academy of Science. 
KW's visit to the University of Hong Kong was provided by the Royal Society Kan Tong Po Professorship award.

\appendix

\bsp

\label{lastpage}


\begin{thebibliography}{99}
\bibitem[\protect\citeauthoryear{Angel}{1969}]{Angel69} Angel J.R.P., 1969, ApJ, 158, 219
\bibitem[\protect\citeauthoryear{Arimoto et al.}{2008}]{Arimoto08} Arimoto M., Tsubuku Y., Toizumi T. et al., 2008, AIP Conf. Proc., 1000, 607  
\bibitem[\protect\citeauthoryear{Begelman \& Sikora}{1987}]{Begelman87} Begelman M.C., Sikora M., 1987, ApJ, 322, 650
\bibitem[\protect\citeauthoryear{Bellazzini et al.}{2007}]{Bellazzini07} Bellazzini R., et al., 2007, Nucl. Instr. Meth. Phys. Res. A., 579, 853
\bibitem[\protect\citeauthoryear{Bjornsson}{1982}]{Bjornsson82a} Bjornsson C.-I., 1982, ApJ, 260, 855
\bibitem[\protect\citeauthoryear{Bjornsson \& Blumenthal}{1982}]{Bjornsson82b} Bjornsson C.-I., Blumenthal G.R., 1982, ApJ, 259, 805
\bibitem[\protect\citeauthoryear{Blandford \& Eichler}{1987}]{Blandford87} Blandford R., Eichler D., 1987, Physics Reports, 154, 1
\bibitem[\protect\citeauthoryear{Bonometto \& Saggion}{1973}]{Bonometto73} Bonometto S., Saggion A., 1973, A\&A, 23, 9
\bibitem[\protect\citeauthoryear{Bregman}{1990}]{Bregman90} Bregman J.N., 1990, Astron. Astrophys. Rev., 2, 125
\bibitem[\protect\citeauthoryear{Celotti \& Ghisellini}{1999}]{Celotti99} Celotti A., Ghisellini G., 1999, proc. Gamma-ray Bursts: the First Three Minutes, ASP, 190, 180
\bibitem[\protect\citeauthoryear{Celotti \& Matt}{1994}]{Celotti94} Celotti A., Matt G., 1994, MNRAS, 268, 451
\bibitem[\protect\citeauthoryear{Celotti, Ghisellini \& Chiaberge}{2001}]{Celotti01} Celotti A., Ghisellini G., Chiaberge M., 2001, MNRAS, 321, L1
\bibitem[\protect\citeauthoryear{Celotti et al.}{1998}]{Celotti98} Celotti A., Kuncic Z., Rees M.J., Wardle J.F.C., 1998, MNRAS, 293, 288
\bibitem[\protect\citeauthoryear{Costa et al.}{2001}]{Costa01} Costa E., Soffitta P., Bellanzzini R., Brez A., Lumb N., Spandre G., 2001, Nature, 411, 662 
\bibitem[\protect\citeauthoryear{Costa et al.}{2008}]{Costa08} Costa E., Bellanzzini R., Bregeon J., et al., 2008, arXiv:0810.2700 
\bibitem[\protect\citeauthoryear{Cullen}{2001a}]{Cullen01a} Cullen J.G., 2001a, PhD Thesis, University of Sydney
\bibitem[\protect\citeauthoryear{Cullen}{2001b}]{Cullen01b} Cullen J.G., 2001b, JCoPh, 173, 175 
\bibitem[\protect\citeauthoryear{Dermer \& Schlickeiser}{1993}]{Dermer93} Dermer C.D., Schlickeiser R., 1993, ApJ, 416, 458
\bibitem[\protect\citeauthoryear{Dolan}{1967}]{Dolan67} Dolan J.F., 1967, Space Science Reviews, 6, 579  
\bibitem[\protect\citeauthoryear{Fender}{2006}]{Fender06} Fender R., 2006, in Lewin H.G., van der Klis M. eds, Compact Stellar X-ray Sources, Cambridge Astrophysics Series 
\bibitem[\protect\citeauthoryear{Freeland et al.}{2006}]{Freeland06} Freeland M., Kuncic Z., Soria R., Bicknell G.V., 2006, MNRAS, 372, 630
\bibitem[\protect\citeauthoryear{Gunji et al.}{2001}]{Gunji04} Gunji S., Suzuki T., Sato F., Sakurai F., Tokanai Y., Saito Y., Kubota A., 2004, Advances in Space Research, 33, 10 
\bibitem[\protect\citeauthoryear{Harris \& Krawczynski}{2006}]{Harris06} Harris D.E., Krawczynski H., 2006, Annual Review of Astronomy and Astrophysics, 44, 463
\bibitem[\protect\citeauthoryear{Heitler}{1936}]{Heitler36} Heitler W., 1936, Quantum Theory of Radiation, Oxford University Press, p. 146
\bibitem[\protect\citeauthoryear{Hua}{1997}]{Hua97} Hua X.-M., 1997, Computers in Physics, 11, 6 
\bibitem[\protect\citeauthoryear{Jahoda et al.}{2007}]{Jahoda07} Jahoda K., Black K., Deines-Jones P., Hill J.E., Kallman T., Strohmayer T., Swank J.H., 2007, arXiv:astro-ph/0701090v1
\bibitem[\protect\citeauthoryear{Jauch \& Rohrlich}{1980}]{Jauch80} Jauch J.M., Rohrlich F., 1980, The Theory of Photons and Electrons, Springer-Verlag, p. 229
\bibitem[\protect\citeauthoryear{Jorstad et al.}{2007}]{Jorstad07} Jorstad S.G. et al., 2007, ApJ, 134, 799
\bibitem[\protect\citeauthoryear{Kataoka \& Stawarz}{2005}]{Kataoka05} Kataoka J., Stawarz L., 2005, ApJ, 622, 797
\bibitem[\protect\citeauthoryear{Kovalev et al.}{2007}]{Kovalev07} Kovalev Y.Y., Lister M.L., Homan D.C., Kellermann K.I., 2007, ApJ, 668, L27
\bibitem[\protect\citeauthoryear{Kuncic, Wu \& Cullen}{2005}]{Kuncic05} Kuncic Z., Wu K., Cullen J.G., 2005, Publ. Astron. Soc. Aust., 22, 56 
\bibitem[\protect\citeauthoryear{Lapidus, Syunyaev \& Titarchuk}{1985}]{Lapidus85} Lapidus I.I., Syunyaev R.A., Titarchuk L.G., 1985, Astrophysics, 23, 663 
\bibitem[\protect\citeauthoryear{Lyutikov, Pariev \& Gabuzda}{Lyutikov et al.}{2005}]{Lyutikov05} Lyutikov M., Pariev V.I., Gabuzda D.C., 2005, MNRAS, 360, 869
\bibitem[\protect\citeauthoryear{Maraschi, Ghisellini \& Celotti}{1992}]{Maraschi92} Maraschi L., Ghisellini G., Celotti A., 1992, ApJ, 397, L5
\bibitem[\protect\citeauthoryear{Marshall et al.}{2005}]{Marshall05} Marshall H.L., Schwartz D.A., Lovell J.E.J., Murphy D.W. et al., 2005, ApJ, 156, 13
\bibitem[\protect\citeauthoryear{Matt}{2004}]{Matt04} Matt G., 2004, A\&A, 423, 495
\bibitem[\protect\citeauthoryear{Matt et al.}{1996}]{Matt96} Matt G., Feroci M., Rapisarda M., Costa E., 1996, Rad. Phys. Chem., 48, 403
\bibitem[\protect\citeauthoryear{McNamara, Kuncic \& Wu}{McNamara et al.}{2008a}]{McNamara08a} McNamara A.L., Kuncic Z., Wu K., 2008a, MNRAS, 386, 2167
\bibitem[\protect\citeauthoryear{McNamara et al.}{2008b}]{McNamara08b} McNamara A.L., Kuncic Z., Wu K., Galloway D.K., Cullen J.G., 2008b, MNRAS, 383, 962
\bibitem[\protect\citeauthoryear{Novick et al.}{1972}]{Novick72} Novick R., Weisskopf M.C., Berthelsdorf R., Linke R., Wolff R.S., 1972, ApJ, 174, L1
\bibitem[\protect\citeauthoryear{Pozdnyakov, Sobol \& Sunyaev}{1983}]{P83} Pozdnyakov L.A., Sobol I.M., Sunyaev R.A., 1983, Ap\&SS Rev. 2, 189
\bibitem[\protect\citeauthoryear{Poutanen}{1994}]{Poutanen94} Poutanen J., 1994, ApJS, 92, 607
\bibitem[\protect\citeauthoryear{Sambruna et al.}{2004}]{Sambruna04} Sambruna R., Gambill J.K., Maraschi L., Tavecchio F., Cerutti R., Cheung C.C., Megan Urry C., Chartas G., 2004, ApJ, 608, 698
\bibitem[\protect\citeauthoryear{Schwartz}{2002}]{Schwartz02} Schwartz, D.A., 2002, ApJ, 569, L23
\bibitem[\protect\citeauthoryear{Shakura \& Sunyaev}{1973}]{SS73} Shakura N.I., Syunyaev R.A., 1973, A\&A, 24, 337 
\bibitem[\protect\citeauthoryear{Sikora, Begelman \& Rees}{1994}]{Sikora94} Sikora M., Begelman M.C., Rees M.J., 1994, ApJ, 421, 153
\bibitem[\protect\citeauthoryear{Sunyaev \& Titarchuk}{1985}]{ST85} Sunyaev R.A. Titarchuk L.G., 1985, A\&A, 143, 374 
\bibitem[\protect\citeauthoryear{Tavecchio}{2007}]{Tavecchio07} Tavecchio F., 2007, Ap\&SS, 311, 247
\bibitem[\protect\citeauthoryear{Tavecchio et al.}{2000}]{Tavecchio00} Tavecchio F., Maraschi L., Sambruna R.M., Urry C.M., 2000, ApJ, 544, L23
\bibitem[\protect\citeauthoryear{Wagner}{1995}]{Wagner95} Wagner S.J., 1995, A\&A, 298, 688
\bibitem[\protect\citeauthoryear{Weisskopf et al.}{2006}]{Weisskopf06} Weisskopf M.C., Elsner R.F., Hanna D., Kaspi V.M., O'Dell S.L., Pavlov G.G., Ramsay B.D., 2006, arXiv:astro-ph/0611483v1
\bibitem[\protect\citeauthoryear{Weisskopf et al.}{1972}]{Weisskopf72} Weisskopf M.C., Berthelsdorf R., Epstein G., Linke R., Mitchell D., Novick R., Wolff R.S., 1972, Rev. Sci. Instr., 43, 967
\end{thebibliography}
\end{document}